\documentclass[preprint, 12pt]{elsarticle}

\newcommand{\figura}[4]{\begin{figure}[ht]

                        {center}
                           \includegraphics[width=#1 cm, height=#2 cm]{#3}
                           \caption{#4}
                           \end{center}
                           \end{figure}}
\usepackage{graphicx}
\usepackage{dcolumn}
\usepackage{bm}
\usepackage{savesym}
\usepackage{amsmath, amsfonts, amsthm, amssymb}
\savesymbol{iint}
\usepackage{txfonts} 
\restoresymbol{TXF}{iint}
\usepackage[ansinew]{inputenc}
\usepackage{textcomp}
\usepackage{comment}
\usepackage{amssymb}
\usepackage{enumerate}

\journal{Astroparticle Physics}
\begin{document}

\begin{frontmatter}



\title{Secluded WIMPs, dark QED with massive photons, and the galactic center gamma-ray excess}

\author{E. C. F. S. Fortes}  
\address{Instituto de F\'\i sica Te\'orica, Universidade Estadual Paulista, Rua Dr. Bento Teobaldo Ferraz 271 01140-070 S\~ao Paulo, SP, Brazil}

\author{V. Pleitez}
\address{Instituto de F\'\i sica Te\'orica, Universidade Estadual Paulista, Rua Dr. Bento Teobaldo Ferraz 271 01140-070 S\~ao Paulo, SP, Brazil}

\author{F. W. Stecker}
\address{Astrophysics Science Division, NASA Goddard Space Flight Center, Greenbelt, MD 20771} 

\vspace{-1.5cm}
\begin{abstract}
We discuss a particular secluded WIMP dark matter model consisting of neutral fermions as the dark matter candidate and a Proca-Wentzel (PW) field as a mediator.  In the model that we consider here, dark matter WIMPs interact with standard model (SM) particles only through the PW field of $\sim$ MeV -- multi-GeV mass particles. The interactions occur {\it via} a $U(1)^{\prime}$ mediator, $V_{\mu}^{\prime}$, which couples to the SM by kinetic mixing with $U(1)$ hypercharge bosons, $B_{\mu}^{\prime}$. One important difference between our model and other such models in the literature is the absence of an extra singlet scalar, so that the parameter with dimension of mass $M^2_V$ is not related to a spontaneous symmetry breaking. This QED based model is also renormalizable. The mass scale of the mediator and the absence of the singlet scalar can lead to interesting astrophysical signatures. The dominant annihilation channels are different from those usually considered in previous work. We show that the GeV-energy $\gamma$-ray excess in the galactic center region, as derived from {\it Fermi}-LAT Gamma-ray Space Telescope data, can be attributed to such secluded dark matter WIMPs, given parameters of the model that are consistent with both the cosmological dark matter density and the upper limits on WIMP spin-independent elastic scattering. Secluded WIMP models are also consistent with suggested upper limits on
a DM contribution to the cosmic-ray antiproton flux.
\end{abstract}
\end{frontmatter}

\maketitle

\section{Introduction}
\label{sec:intro}

The clear astronomical ~\cite{Burstein:1985} and cosmological evidence for large amounts of dark matter (DM) in the universe~\cite{Hinshaw:2013} has led to the construction of various theoretical models that go beyond the standard model (SM) weak-scale theories and which attempt to account for the DM abundance in the universe~\cite{Feng:2010}. The observational evidence for DM has motivated various experimental searches to find dark matter~\cite{Pierce:2008zza}.

Data from colliders are used to search for evidence of dark matter particles. Experiments with detectors like DAMA, CoGeNT, CDMS, XENON and LUX are used to search for evidence of the recoil energy of nuclei that would be produced by scattering with dark matter particles~\cite{Bernabei:2010mq, Aprile:2012}. High-energy colliders like LHC (Large Hadron Collider), have obtained significant upper limits on the annihilation of WIMPs to quarks~\cite{Fox:2012}. They also offer very interesting possibilities to investigate interactions involving DM mediators.

Space-borne detectors have been used to search for evidence of the products of of dark matter annihilation, particularly $\gamma$-rays and cosmic-ray positrons. 
These searches have conservatively produced constraints on dark matter annihilation, both from cosmic $\gamma$-ray studies~\cite{Ackermann:2011} and cosmic-ray positron studies~\cite{Aguilar:2014}. However, analyses of the {\it Fermi}-LAT data have indicated the existence of an "excess" flux of $\gamma$-rays above that expected from cosmic rays interacting with interstellar gas. This flux appears to be extended around the region of the galactic center. It appears to peak in the 2 -- 3 GeV energy range. This excess has been interpreted to be a possible indication of the annihilation of weakly interacting dark matter (WIMP) particles having a mass in the 20 -- 45 GeV range, annihilating primarily into quarks or, less likely, in the 7 -- 12 GeV mass range annihilating primarily into charged $\tau$ leptons~\cite{Hooper:2010mq, Daylan:2014rsa}. Models involving other annihilation channels have also been considered~\cite{Achterberg}. We note, however, the determination of various possible components of $\gamma$-ray emission from the galactic center is complicated and the $\gamma$-ray data are not precise enough to point to a unique origin. Other possible interpretations of this excess that include other contributions to the $\gamma$-ray flux in the region of the galactic center have been suggested~\cite{Abazajian:2011}.

Neutralino supersymmetric WIMPs, {\it viz.}, the lightest supersymmetric dark matter particles, have been a popular choice to be the DM WIMPs because they are stable and neutral and their cross section naturally leads to the correct cosmological DM density.
However, as of now, the LHC has not found any evidence for such particles. Therefore, other candidate WIMP models have been explored and should be further explored.

In this work we consider a model of WIMP dark matter in which the dark sector is just quantum electrodynamics (QED) extended with a new massive photon field, usually dubbed a Proca-Wentzel (PW) field. It is well known that this is a renormalizable theory because it couples to a conserved vector current. Hence, the dark sector is made up of Dirac fermions, $\eta$, that only interact {\it via} a PW field, here denoted by $V_{\mu}^{\prime}$, that serves to mediate between DM fermions and standard model particles.

Diagonalization of the kinetic terms in the Lagrangian give both the standard $Z$ boson and an extra neutral gauge boson that we denote as $Z^{\prime}$. We will show that this DM model can produce the observed cosmological DM abundance and that its annihilation into standard model (SM) particles can also lead to astronomically observable fluxes of $\gamma$-rays in the galactic center and in dwarf galaxies.

In our model secluded DM interactions occur only through $Z^{\prime}$ mediators which subsequently decay to SM particles. Thus, they have an very small elastic scattering cross section with nuclei. This distinguishes such secluded WIMP models from other WIMP models. The decay of $Z^{\prime}$'s into quark-antiquark channels produces pions, among which are $\pi^0$'s that decay to produce $\gamma$-rays. The $\pi^0$-decay $\gamma$-ray spectrum has a characteristic peak at $m_{\pi}/2$~\cite{Stecker:1971} and is bounded by the rest-mass of the WIMP, $m_{\eta}$ ({\it e.g,},~\cite{Stecker:1989}). The $Z^{\prime}$ decay, particularly into charged leptons and light quarks, also yields $\gamma$-rays through internal bremsstrahlung. In this process, a $\gamma$-ray spectrum is produced that peaks near $m_{\eta}$~\cite{Bringmann:2008}. Electrons resulting from this process can produce $\gamma$-rays {\it via} Compton scattering in the interstellar medium.

Other $Z^{\prime}$ models have recently been discussed in the context of astrophysical $\gamma$-ray production (see, {\it e.g.}, Ref.~\cite{Hooper:2015}). However, our model, in which the $Z^{\prime}$ is the mediator of the DM interactions through the annihilation channel $\eta + \overline{\eta} \rightarrow Z^{\prime} + Z^{\prime}$, as shown in Figure 1, and with $m_{Z^{\prime}} \ll m_{\eta}$, was not considered in Ref.~\cite{Hooper:2015}. (See also the discussion in Section \ref{sec:differences}).

The outline of the paper is as follows: In section \ref{sec:model} we present the model. In section \ref{sec:differences} we discuss the differences between our model and previously explored WIMP DM models. In section \ref{density} we calculate the relic density of our Dirac fermion DM candidate. In section \ref{ann} we discuss the astrophysical production of $\gamma$-rays from from annihilation of our secluded WIMPs as a possible explanation of the so-called $\gamma$-ray excess from the galactic center and potential $\gamma$-ray signals from the Milky Way satellite dwarf galaxies. In section \ref{concl} we summarize our results and conclusions. In appendix \ref{sec:int} we give the full couplings for the interactions described by the model.

\section{The Basic Model}
\label{sec:model}

The possible existence of an extra $U(1)^{\prime}$ symmetry of nature beyond the SM has been considered for a long time. The addition of this new symmetry factor to the electroweak $SU(2) \otimes U(1)$ of the SM occurs {\it via} the so-called kinetic mixing portal, by mixing with the hypercharge gauge boson $B^{\mu}$. A real massive vector field coupled to a fermionic vector current is a well behaved theory. The mass term breaks what would-be  a gauge symmetry (which is valid in the kinetic term). However, this only implies the constraint $\partial_\mu V^{\prime \mu}=0$. For earlier references see~\cite{Galison:1983pa}.

The dark matter fermion WIMPs of the model, $\eta$, interact only with the PW field that is the connection between DM and ordinary matter, designated by $V_{\mu}^{\prime}$. The PW field mixes through the kinetic term with the $U(1)_Y$ vector field of the SM, at this stage designated by $B^\prime_\mu$. DM interactions occur only through $Z^{\prime}$ mediators which subsequently decay to SM particles. As such, secluded DM WIMPs have a very small elastic scattering cross section with nuclei. This distinguishes secluded WIMP models from other WIMP models, thus allowing for a potential experimental test.

After the diagonalization of the mixing in the kinetic terms,
\begin{equation}
V_{\mu}^{\prime} =  \left(1/\sqrt{1-g_{VB}^2} \ \right) V_{\mu},
\end{equation}
where $V_{\mu}$ is a linear combination of $Z$ and the extra neutral gauge boson $Z^{\prime}$. Cosmological DM abundance constraints on the DM annihilation cross section will then require a small mixing angle between $V$ and $Z^{\prime}$ so that $V \simeq Z^{\prime}$~\cite{Pospelov:2007mp}.  Depending on the mass of $Z^{\prime}$, interesting signatures for these types of mediators could come from the  Drell-Yan channel $pp\rightarrow Z^{\prime}\rightarrow l\overline{l}$ and from non-conventional decays of SM Higgs boson such as $h \rightarrow Z Z^{\prime}$, potentially observable with the Large Hadron Collider (LHC). In this work we consider a scenario where the $Z^{\prime}$ is light, such that $M_{Z^{\prime}} \ll m_{\eta}$. The scalar sector is the same as that of the SM, viz., a doublet with $Y=+1$.

Empirical bounds on $Z^{\prime}$ couplings exist for these kinds of models. Such bounds depend on the mass scale of $Z^{\prime}$. High energy colliders are sensitive to $M_{Z^{\prime}}\gtrsim 10$ GeV and the constraints for lighter $Z^{\prime}$'s  are given mainly by precision QED observables, $B$ meson decay and some fixed target experiments~\cite{Curtin:2014cca}.
Besides these constraints, there is also a constraint on the lifetime of a $Z^{\prime}$. Its lifetime should be less than one second in order to guarantee that the $Z^{\prime}$ decays before the onset of big-bang nucleosynthesis~\cite{Serpico:2004nm}. As a consequence, the secluded DM fermion will annihilate preferably into a $Z^{\prime}$ pair~\cite{Pospelov:2007mp}.

The full Lagrangian of the model is given by

\begin{equation}
\mathcal{L}=\mathcal{L}_{SM}+\mathcal{L}_{Dark}+\mathcal{L}_{Dark+int},
\label{e1}
\end{equation}
where $\mathcal{L}_{SM}$ is the SM Lagrangian, and $\mathcal{L}_{Dark}$ is the dark  Lagrangian given by
\begin{equation}
\mathcal{L}_{Dark}=\overline{\eta}(i \partial_\mu \gamma^{\mu}\eta-m_{\eta})\eta.
\label{e2}
\end{equation}
The mixing between dark and SM matter, occurs through the third term in Eq.~(\ref{e1}), which is
\begin{eqnarray}\label{e3}
  \mathcal{L^{\prime}}_{Dark+int}&=&g_{\eta}\overline{\eta}\gamma^{\mu}\eta V^{\prime}_{\mu}+\frac{1}{2}M_{V}^{2}V^{\prime}_{\mu}V^{\prime \mu} \\\nonumber
   && -\frac{1}{4}V^{\prime}_{\mu\nu}V^{\prime\mu\nu}+\frac{g_{VB}}{2}V^{\prime}_{\mu\nu}B^{\prime\mu\nu}-\frac{1}{4}B^{\prime}_{\mu\nu}B^{\prime\mu\nu},
\end{eqnarray}
where $X_{\mu\nu}=\partial_\mu \nu+\partial_\nu X_\mu$, with $X_\mu=V^\prime_\mu,B^\prime_\mu$,
and $B^\prime_\mu$ is the abelian gauge boson of the $U(1)_Y$ factor in the SM Lagrangian.
In Eqs.~(\ref{e2}) and (\ref{e3}), the masses $m_\eta,M_V$ and the couplings $g_{VB}$ and $g_\eta$ are free parameters.
The mixing in the kinetic term in Eq.~(\ref{e3}) can be diagonalized by a $GL(2,R)$ transformation~\cite{Gopalakrishna:2008dv}
\begin{equation}
\left(\begin{array}{c}
V\\B \end{array}\right)=\left(
\begin{array}{cc}
\sqrt{1-g^2_{VB}}&0\\
-g_{VB} &1
\end{array}\right)
\left(\begin{array}{c}
V^\prime\\ B^\prime
\end{array}
\right)
\label{gl2r}
\end{equation}

Using (\ref{gl2r}) in (\ref{e3}) we obtain

\begin{eqnarray}
  \mathcal{L^{\prime}}_{Dark+int}&=&\frac{g_{\eta}}{\sqrt{1-g_{VB}^2}}\overline{\eta}\gamma^{\mu}\eta V_{\mu}+\frac{M_{V}^{2}}{2(1-g_{VB}^2)}V_{\mu}V^{ \mu} \\\nonumber
   && -\frac{1}{4}V_{\mu\nu}V^{\mu\nu}-\frac{1}{4}B_{\mu\nu}B^{\mu\nu}.
\end{eqnarray}

After
 symmetry breaking we get the mass matrix in the ($B^{\mu}$, $W^{3\mu}$ and $V^{\mu}$) basis

\begin{eqnarray}
\emph{M}^{2}=\frac{g^{2}{\rm v_{h}}^{2}}{4 c_{W}^{2}}U^{\dagger}\left(
        \begin{array}{ccc}
          0 & 0 & 0 \\
          0 & 1 & -\xi s_{W} \\
          0 & -\xi s_{W} & \xi^{2}s_{W}^{2}+4 r \\
        \end{array}
      \right)U,
      \label{gzzp1}
\end{eqnarray}
where $s_W$ is the usual weak mixing angle, $\xi=g_{VB}/\sqrt{1-g^2_{VB}}$, $ r=M^2_V/M^2_Z$, and
where the matrix $U$ is given by

\begin{eqnarray}
  U=\left(
      \begin{array}{ccc}
        c_{W} & s_{W} & 0 \\
        -s_{W} & c_{W} & 0 \\
        0 & 0 & 1 \\
      \end{array}
    \right)
\end{eqnarray}
We can write the $3\times 3$ mass eigenstates matrix as

\begin{equation}
\left(
\begin{array}{c}
B\\ W_3\\ V\end{array}
\right)=\left(\begin{array}{ccc}
c_W&-s_Wc_\alpha&s_Ws_\alpha\\
s_W&c_Wc_\alpha&-c_Ws_\alpha\\
0&s_\alpha&c_\alpha
\end{array}
\right)
\left(
\begin{array}{c}
A\\ Z \\ Z^\prime\end{array}
\right)
\label{gzzp}
\end{equation}

with
\begin{equation}
t_{2\alpha}=-\frac{2s_W\xi}{1-s^2_W\xi^2-r },
\label{alpha}
\end{equation}
where $t_{2\alpha}$ denotes tan~$2\theta_{\alpha}$, $t_{\alpha}$ denotes tan~$\theta_{\alpha}$, $c_{\alpha}$ denotes cos~$\theta_{\alpha}$ and $s_{\alpha}$ denotes the sin~$\theta_{\alpha}$. For small values of $\theta_{\alpha}$, we can expand $t_{\alpha} \simeq t_{2\alpha}/2 - t_{2\alpha}^{3}/8$ and use the relation $c_{\alpha} = (1+t_{\alpha}^{2})^{-1/2}$ to perform the calculations.

The masses of the $Z$ and $Z^{\prime}$ are found by diagonalizing the matrix (\ref{gzzp1}). They are given by:

\begin{equation}\label{mzzp}
    M_{Z,Z^{\prime}}=\frac{M_{Z_{0}}^{2}}{2}[(1+s_{W}^2\xi^2+ r)\pm \sqrt{(1-s_{W}^2\xi^2- r)^{2}+4s_{W}^2\xi^2}],
  \end{equation}
\noindent where $M_{Z_{0}}=g {\rm{v_h}}/(2 c_{W})$, assuming that $\xi\ll 1$ and $r< (1-s_{W}^2\xi^2)$. (Again, we assume that the lighter neutral vector boson is $Z^ \prime$). Notice that for the  electroweak precision observables, it follows that as we increase $M_{Z^{\prime}}$ we have to decrease the coupling $g_{VB}$ in order to respect the  constraint which implies that $\xi/\sqrt{|1-M_{Z}^{\prime 2}/M_{Z}^{2}|}\lesssim 10^{-2}$.

\section{Classification of Heavy Photon Models}
\label{sec:classification}

Electroweak models with an extra $U(1)$ symmetry are among most well motivated extensions of the Standard Model. In a general context we use the notation
$U(1)_1\otimes U(1)_2$. All these models allow a kinetic mixing between the field strength tensors of both $U(1)$ gauge bosons, $F_{1\mu\nu}F_2^{\mu\nu}$.

We can separate this sort of models in several groups. For instance,
\begin{enumerate}
\item[A1] Both $U(1)$ groups are visables. It means that SM fermions carry both quantum numbers. This possibility usually arises in the context of grand unified theories (GUTs). Examples are the models of Babu et al. and that of Galison and Manohar~\cite{Galison:1983pa}. Models in which $U(1)_1 = U(1)_y$ and $U(1)_2 = U(1)_{B-L}$ are of this type.
\item[A2] There are fermions which carry only one of the $U(1)$ charges, others carry both charges and some with no charges. As an example of these sort of models we have del Aguila et al. in ~\cite{Galison:1983pa} .
\item[A3] One of the $U(1)$ factor is visible, the SM particles carry one of the $U(1)$ charge, for example $U(1)_1$, an the other, $U(1)_2$ is dark. Our model is of this type.
\end{enumerate}

Another way to classify these sort of models is by considering the mass scale at which  the kinetic mixing occurs. Although it is not usually explicitly say, this is an important point.
\begin{enumerate}
\item[B1] Models in which $U(1)_1=U(1)_Y$. The kinetic mixing occurs  \textit{before} the SSB. Our model and the one of Ref.~\cite{Gopalakrishna:2008dv} are of this type.
\item[B2] Models in which $U(1)_2=U(1)_Q$, i.e., the kinetic mixing occurs \textit{after} the SSB and the mixing is with the massless photon. An example of this type of model is that of Holdom in~\cite{Galison:1983pa}.
\item[B3] Neither $U(1)_1$ nor $U(1)_2$ are related to the symmetry of the SM directly, but $U(1)_1\otimes U(1)_2\to U(1)_Y$ after SSB. The Galison and Manohar model~\cite{Galison:1983pa} is of this type.
\end{enumerate}

There are also models in which
\begin{enumerate}
\item[C1] The mixing among the two $U(1)$ factor occur in  the kinetic and also in the mass terms. In this case the photon has a component on $Z^\prime$. See, for instance, the model of Babu \textsl{et al.}, in Ref.~\cite{Galison:1983pa}.
\item[C2] The mixing among the two $U(1)$ factors occurs only in the kinetic term. Examples of these sort of model are our model and that of Ref.~\cite{Gopalakrishna:2008dv}. In these models the photon has no component on $Z^ \prime$.
\end{enumerate}
We  can also classify the models according to
\begin{enumerate}
\item[D1] Extra scalars, for instance singlets, are added to break the additional $U(1)$ symmetry.
\item[D2] The only scalar in the model is that of the SM. Our model is of this type. 	
\end{enumerate}

\noindent Models may also be classified according to the fermion, scalar or vector nature of both the DM and the mediator.
\begin{enumerate}
\item[E1] Our model has a Dirac fermion as DM and a real massive vector as mediator.
The model of Ref.~\cite{Liu:2014} postulates a real scalar or a vector field to be the dark force and a complex scalar or a fermion to be the dark matter.
\item[F1] The St\"{u}ckelberg scalar model was introduced in order to maintain the gauge invariance in a QED-like theory with a massive real vector field. In this sense it provides an alternative to the Higgs mechanism. However, it needs an axionic scalar $S$ ($\phi$ in the notation of Ref.~\cite{Kors:2005uz}). 
\end{enumerate}

\section{Specific differences between other models and our model}
\label{sec:differences}

Our model is motivated by the following consideration. It is well known that massive quantum electrodynamic is a renormalizable and free of anomalies theory. It means that spontaneous symmetry breaking in order to give mass to the vector field can be implemented, but it is not mandatory. Our model has a Dirac fermion as DM and a real massive vector as mediator.

The important point is that the photon (massless or not) couples to a conserved current. This occurs because the bad high energy behavior of the vector propagator $\sim k_\mu k_\nu/m^2_V$ vanishes when contracted with the conserved current, say $J^\mu$. The $U(1)^\prime$ symmetry in the kinetics term is broken by the mass term $(m^2_V/2)V_\mu V^\mu$, but its only consequence is the constraint $\partial_\mu V^\mu = 0$, which comes from the equation of motion.
It is still possible to restore the gauge invariance if a massive scalar field called a
St\"{u}ckelberg field is invoked. If $S(x)$ denotes the gauge St\"{u}ckelberg scalar, satisfying the equation $(\Box -m^2_V)S(x) = 0$.

We note that the Proca-Wentzel Lagrangian that we use is, in fact, equivalent to the St\"{u}ckelberg Lagrangian when $S(x)=0$ (For the main references, see~\cite{Ruegg:2003ps}). Thus, our model coincides with that of Ref. \cite{Kors:2005uz} (F1 above) when no St\"{u}ckelberg field is present. An important difference between our model and a St\"{u}ckelberg-like model is that the later predicts the existence of millicharged particles~\cite{Feldman:2007wj}. That model introduces a new unit of electric charge, allowing the electric charge to have non-integer values. Hence, although both models have similar features there are many differences between the models.

In our model we have a Dirac fermion and a vector mediator and the mediator is lighter than the DM candidate. This is also the case in Ref.~\cite{Hooper:2012}.  However, in the later model, the leading interaction between the Standard Model and the dark sector is the kinetic mixing between the photon and the $U^\prime(1)$ gauge boson: $\mathcal{L}=(1/2)\epsilon F^\prime_{\mu\nu}F^{\mu\nu}$. The leading kinetic mixing with the photon  is also used  in  Ref.~\cite{Davoudiasl:2015hxa}.

\section{Details of the model}
\label{details}

In the present model, the kinetic mixing is with the gauge boson of the $U(1)_Y$ factor \textit{before} the spontaneous symmetry breaking induced by the Higgs boson in the SM that we have denoted $B^ \prime$. It is in this state that the mixing in Eq.~(5) occurs. Hence, the leading interaction between the SM particles  and the dark sector is through the $W^ 3$ and $B$ component on $Z^ \prime$. See equation (9).
It is through the mixing with the $Z$ boson that the particles in the SM have the couplings with $Z^ \prime$ given in the Appendix.

A model in which the photon has no dark component is that in Ref.~\cite{Gopalakrishna:2008dv}, however,  they introduce an scalar singlet $\Phi_H$, to break the  $U(1)^\prime$ spontaneously. We denote the vacuum expectation value of this singlet ${\rm v_S}$, as opposed to the VEV of the SM Higgs, which we denote by ${\rm v_h}$ . The scalar singlet make the difference of the model in Ref.~\cite{Gopalakrishna:2008dv} and ours. In former case the interactions the of Higgs bosons are  
\begin{eqnarray} \nonumber
h ff: -ic_h\frac{m_{f}}{{\rm {v_h}}},  \nonumber \\
h WW: 2ic_h\frac{m_{W}^2}{{\rm {v_h}}},  \nonumber \\
h Z Z: 2ic_h\frac{m_{Z}^2}{{\rm {v_h}}}(c_{\alpha}-\xi s_{W}s_{\alpha})^2-is_h
\frac{m^2_V}{{{\rm {v}_S}}}{s^2_{\alpha}},  \nonumber \\
h Z^{\prime}Z^{\prime}: 2ic_h\frac{m_{Z}^2}{{\rm {v_h}}}(s_{\alpha}+\xi s_{W}c_{\alpha})^2-is_h\frac{m^2_V}{\textrm{v}_S}c^2_\alpha, \nonumber \\
h Z Z^{\prime}: 2ic_h\frac{m_{Z}^2}{{\rm {v_h}}}[(c_{\alpha}-s_{W}\xi s_{\alpha})(s_{\alpha}+s_{W}\xi c_{\alpha})]-is_h\frac{m^2_V}{\textrm{v}_S}s_\alpha c_\alpha. \nonumber
\label{333}
\end{eqnarray}
where  $\textrm{v}_S$ is the VEV of the scalar singlet (designated by $\xi$ in Ref.~\cite{Gopalakrishna:2008dv}) that is needed to break the $U(1)^\prime$ symmetry and $c_h=\cos\theta_h,s_h=\sin\theta_h$ are the cosine and sine of the mixing angle:
\begin{equation}
\left( \begin{array}{c}
\phi_{SM}\\ \phi_H
\end{array}
\right)=\left(
\begin{array}{cc}
c_h & s_h\\
-s_h &c_h
\end{array}
\right)\left(
\begin{array}{c}
h \\H
\end{array}
\right).
\label{ch}
\end{equation}
with $h$ and $H$ being the mass eigenstates. The scalar singlet carries $U(1)^\prime$  charge denoted  $e^\prime$ in~\cite{Pospelov:2007mp}.

Moreover, the scalar potential has several dimensionless parameters in the quartic interaction terms
\begin{equation}
-V^{(4)}(\Phi,S)=\lambda (\Phi^\dagger_{SM}\Phi_{SM})^2+\rho (S^\dagger S)^2+\kappa S^\dagger \Phi^\dagger_{SM}\Phi_{SM}
\label{potential}
\end{equation}
which obeys the relation~\cite{Gopalakrishna:2008dv}.
\begin{equation}
\label{tan2th}
\tan (2\theta_h) = {\frac{\kappa {\rm v_h}{\rm {v}_S}}{\sqrt{\rho {\rm {v_h}}^2_S - \lambda{\rm {v_h}}^2}}}. 
\end{equation}

In a model with a scalar singlet, the SM-like Higgs scalar has weaker interactions with fermion and vector bosons. Such interactions are suppressed by a factor of $c_h$ relative to SM interactions. The vertices $hZZ$, $hZZ^\prime$ and $hZ^\prime Z^\prime$ have an extra term proportional to $s_h$. Only when $c_h = 1$, do the interactions with the SM particles for both models agree. However, in the models of Refs.~\cite{Pospelov:2007mp} and~\cite{Gopalakrishna:2008dv},
the dark vector still  interacts with the singlet $h$. In general, the phenomenology of both models is a bit different in processes like $h\to Z^\prime Z,Z^\prime Z^\prime,ZZ$ ~\cite{Curtin:2014cca}.  This may imply that the region allowed for the four parameters $m_\eta, g_\eta, g_{VB}$ and $m_V$ is different in both models, although an overlap may  also exist, as follows from the DM calculation above. For example, in the models of Refs.~\cite{Pospelov:2007mp} and~\cite{Gopalakrishna:2008dv},
the mass of the vector field $m_V=g_\eta \textrm{v}_S$ and the mass of $h$ also depends on $v_S$.  Hence, constraints on $m_V$ are also constraint on $m_h$ which could be in disagreement with accelerator data. This relation does not exist in our model in which there is no scalar singlet. In our model $m_V$ is a free parameter. These interactions are equal to those in the appendix only when $c_h = 1$ in equation~(\ref{333}).

Notice that we are considering the $Z^\prime$ to be lighter than the $Z$, then in our Eq.~(11) the signal $-$ corresponds to the lighter boson. In fact, it may have zero mass if $M_V=0$. In this case we have a dark QED indeed. We also note that the values that we take for $g_{VB}$
(called $\epsilon$ in experimental papers) are much lower than the present experimental upper
limits~\cite{Lees:2014} but may be within the sensitivity range of future experiments~\cite{Izaguirre:2015}.

\begin{figure}
  \centering
 \includegraphics[width=7.0cm]{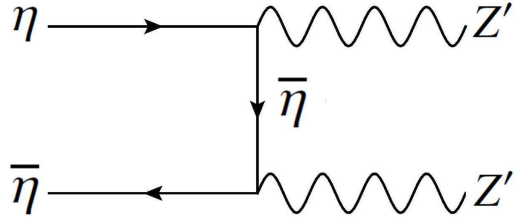}\\
  \caption{Feynman diagram for $\eta + \overline{\eta} \rightarrow Z^{\prime} + Z^{\prime}$.}\label{figure1}
\end{figure}

\section{Cosmological DM Density}
\label{density}

${}$
As noted before, in this model, in addition to SM particles, there is a dark sector composed of a Dirac fermion, $\eta$, and a Proca-Wentzel vector field, $V_\mu \simeq Z^{\prime}$, in the limit of small mixing angle. We consider $\eta$ to be the only component of DM.  In the regime where $m_{\eta}\gg M_{Z^{\prime}}$, the primary annihilation channel will be $\eta+\overline{\eta}\rightarrow Z^{\prime}+ Z^{\prime}$~\cite{Pospelov:2007mp}, as shown in Figure \ref{figure1}. The differential annihilation cross section in this case is given by
\begin{eqnarray}\label{difcross}
  \frac{d\sigma_{ann}}{d\Omega^{\prime}} &=& \frac{c_{\alpha}^{4}g_{\eta}^{4}}{64\pi s(1-g_{VB}^{2})}\sqrt{\frac{s}{s-4m_{\eta}^{2}}}\times \\\nonumber
  && \frac{[A_{1}+\cos\theta[-A_{2}^{3}+ A_{2}^{2}\cos\theta(4m_{\eta}^{2}+s+A_{2}\cos\theta)]] }{s(A_{2}\cos\theta-s)^{2}(s+A_{2}\cos\theta)},
\end{eqnarray}
where $d\Omega^{\prime} =  d\varphi \sin\theta d\theta$, $A_{1}=s(32m_{\eta}^{4}-8m_{\eta}^{2}s - s^{2})$ and $A_{2}=\sqrt{s(s-4m_{\eta}^{2})}$, where the summation over final spins and average over initial spins are taken into account.

After integrating equation (\ref{difcross}) to obtain the total annihilation cross section, we used the approximation for the square of the center-of-momentum energy, $s \simeq 4m_{\eta}^{2}+m_{\eta}^{2}v^{2}$, which is valid for non-relativistic particles.  The annihilation cross-section can then be expressed in the form $\langle\sigma_{ann}\mid v\mid \rangle= a + b v^{2}$, where the $a$ and $b$ are given in equation (\ref{sigv}).  Note that relation between the velocity in the center-of-momentum ($v_{cms}$) and the relative velocity ($v$) is given by $v_{cms} = v/2$ ~\cite{Wells:1994qy}. We then find
\begin{equation}\label{sigv}
   \langle\sigma_{ann} v\rangle \simeq \frac{c_{\alpha}^{4}g_{\eta}^{4}}{32\pi (1-g_{VB}^2)m_{\eta}^2}+\frac{3}{8}\frac{c_{\alpha}^{4}g_{\eta}^{4}}{32\pi (1-g_{VB}^2)m_{\eta}^2}v^{2}. 
\end{equation}
Using this expression, we calculated relic density as a numerical solution to the Boltzmann equation, discussed in~\cite{Beltran:2008xg, Lee:2007mt}
\begin{equation}
\label{omegas}
  \Omega h^{2} \simeq \frac{1.04\times 10^{9} \ {\rm GeV^{-1}}}{M_{pl}}\frac{X_{f}}{\sqrt{g_{*}(X_{f})}}\frac{1}{a+3b/X_{f}},
\end{equation}
where $X_{f}$ is given by
\begin{equation}\label{xf}
  X_{f}=\ln\left[c(c+2)\sqrt{\frac{45}{8}}\frac{g m_{\eta}M_{pl}(a+6b/X_{f})}{2\pi^{3}\sqrt{g_{*}(X_{f})}}  \right],
\end{equation}
and where $g=2$ for fermionic DM, the Planck max, $M_{pl} = 1.22\times 10^{19}$ GeV, $c$ is a parameter of order unity considered here as 5/4 and $g_{*}(X_{f})$ is the number of relativistic degrees of freedom at freeze-out. Given one neutral Dirac massive particle and one additional neutral gauge boson and all the SM content, $g_{*}(X_{f})\approx 113.25$, $X_{f}=m_{\eta}/T_{f}$ and $T_{f}$ is the temperature at freeze-out. For relics with mass in the range of electroweak scale, $X_{f}$ is in the range 20 -- 30.

Another parameterization for $\langle\sigma_{ann} v\rangle$, given in Refs.~\cite{Pospelov:2007mp} and~\cite{Pospelov:2008jd}, is often used to obtain the annihilation cross section for fermionic dark matter. In the case where the mass of the mediator is negligible, since $g_{VB} \ll 1$, our equation~(\ref{sigv}) reduces to equation (6) in Ref.~\cite{Pospelov:2008jd}, {\it viz.}, $\langle\sigma_{ann} v\rangle = (1/2){\pi (\alpha'^{2}}/{m_{\eta}^2)}$.\footnote{The factor 1/2 in equation this equation comes from averaging over the $U(1)^{\prime}$ charges.}. It follows from the identification of these two equations that the value of the parameter that determines the correct DM abundance, {\it viz.} $\alpha^{\prime}$, is actually independent of $m_{\eta}$. Considering that the annihilation in the early universe is predominantly s-wave, we can neglect the $b$ term in equation~(\ref{sigv}) in the s-wave approximation and identify 
\begin{equation}
\label{alphaprime}
\alpha^{\prime} \simeq {{c_{\alpha}^2 g_{\eta}^2}\over{4\pi\sqrt{1-g_{VB}^2}}}.
\end{equation}

\begin{figure}
  \centering
  \includegraphics[width=7.7cm]{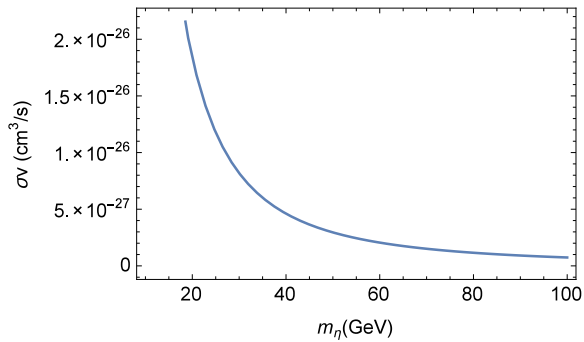}\\
  \caption{$\langle \sigma_{ann} v\rangle$ in $cm^{3}/s$ as a function of $m_{\eta}$ in the limit where $M_{Z^{\prime}}/m_{\eta}\rightarrow 0$, $\xi \sim g_{VB}=8\times 10^{-9}$, $g_{\eta} = 0.088$.  }
  \label{figure2}
\end{figure}

\begin{figure}
  \centering
  \includegraphics[width=7.7cm]{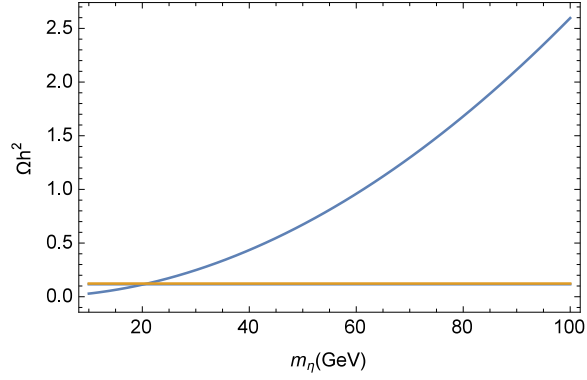}\\
  \caption{ $\Omega h^{2}$ as a function of $m_{\eta}$ in the limit where $M_{Z^{\prime}}/m_{\eta}\rightarrow 0$, $\xi \sim g_{VB}=8\times 10^{-9}$, $g_{\eta} = 0.088$.   The horizontal lines
denote the range of $\Omega h^{2}$ measured by Planck Collaboration~\cite{Planck}. }\label{figure3}
\end{figure}

\begin{figure}
  \centering
  \includegraphics[width=7.7cm]{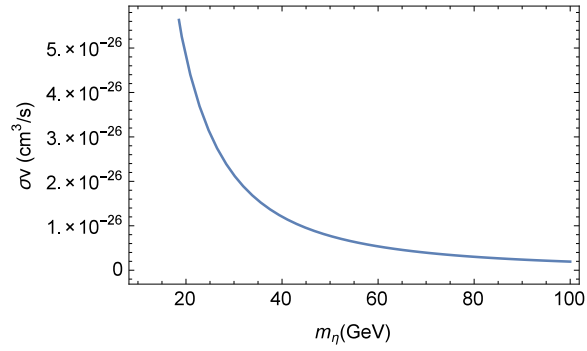}\\
  \caption{$\langle \sigma_{ann} v\rangle$ in $cm^{3}/s$ as a function of $m_{\eta}$,  $\xi \sim g_{VB} = 10^{-6}$, $g_{\eta} = 1.12\times 10^{-1}$, $M_{Z^{\prime}} = 3.5$ GeV.  }
  \label{figure4}
\end{figure}

\begin{figure}
  \centering
  \includegraphics[width=7.7cm]{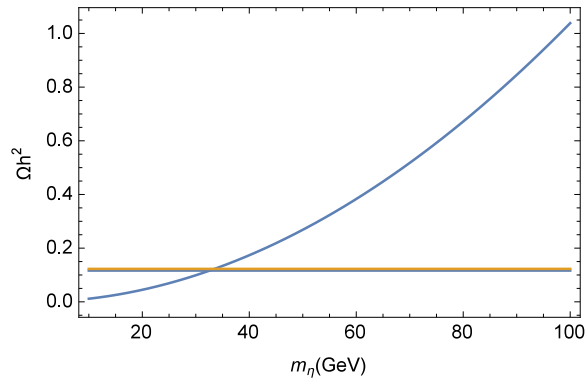}\\
  \caption{$\Omega h^{2}$ as a function of $m_{\eta}$, $\xi \sim g_{VB}=10^{-6}$, $g_{\eta} = 1.12 \times 10^{-1}$, $M_{Z^{\prime}} = 3.5$ GeV. Again using equation (\ref{alpha}). The horizontal lines denote the range of $\Omega h^{2}$ measured by Planck Collaboration~\cite{Planck}.}
  \label{figure5}
\end{figure}

\begin{figure}
  \centering
  \includegraphics[width=7.7cm]{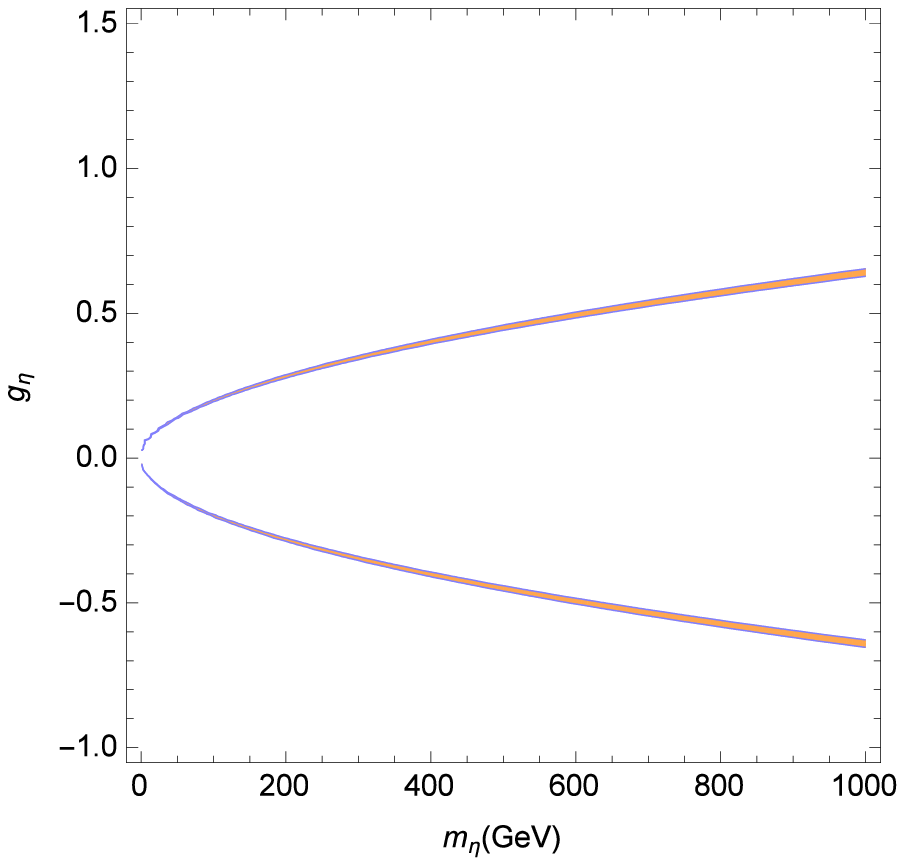}\\
  \caption{Relationship between the parameters $g_{\eta}$ and $m_{\eta}$ giving the correct DM relic abundance within $3\sigma$.}
  \label{figure6}
\end{figure}

Figures \ref{figure2} -- \ref{figure5} give the results for  $<\sigma_{ann} v>$  and $\Omega h^2$
as a function of $m_{\eta}$ in the limit where $M_{Z^{\prime}}/m_{\eta} \rightarrow 0$, taking the values for the model parameters as indicated in the figure captions. Figure~\ref{figure6} shows the relation between $m_{\eta}$ and $g_{\eta}$ which yields the correct DM abundance to within 3$\sigma$. The spread in the curves is determined by taking $\Omega_{DM} = 0.1196$ and a 3$\sigma$ spread with $\sigma = 0.0031$.

\section{Gamma-rays from $\eta$ Annihilation}
\label{ann}

We have pointed out that our particular secluded WIMP dark matter model has certain attractive theoretical features. Having further shown that our proposed DM candidate can yield the proper cosmological relic density, we next consider the production of astrophysical $\gamma$-rays from $\eta$ annihilation.

The $\gamma$-ray flux summed over all possible annihilation channels leading to $\gamma$-ray production is given by
\begin{eqnarray}
  \frac{d\Phi_{\gamma}^{DM}}{d E_{\gamma}} &=& \frac{1}{4\pi m^{2}_{\eta}}\Sigma_{i}\langle \sigma_{i}v\rangle \frac{dN_{\gamma}^{i}}{dE_{\gamma}}\\\nonumber
   &\times& \frac{1}{\Delta\Omega}\int_{\Delta\Omega}d\Omega\int_{l.o.s}\rho^{2}[r(s)]ds,
   \end{eqnarray}\label{fluxo}
where $m_{\eta}$ denotes the mass of the DM candidate and, $\langle\sigma v\rangle $ denotes the thermal averaged annihilation cross section. Since in our model $m_{\eta}>M_{Z^{\prime}}$, the dominant annihilation channel, as already mentioned, will be $\eta+\overline{\eta}\rightarrow Z^{\prime}+Z^{\prime}$ This is then followed by $Z^{\prime}$ decay. Thus, in our case, the resulting $\gamma$-ray spectrum is then the sum of all $Z^{\prime}$ decay channels weighted by their branching ratios and converted from hadrons and muons to $\gamma$-rays.  The integrals in equation (\ref{fluxo}) are integrals over the line of sight to the target and averaged over the solid angle of the region of interest, $\Delta\Omega$.

  Various profiles of the radial distribution of dark matter in the galaxy have been put forth, see e.g., \cite{Cirelli:2010xx}, \cite{Zhao:1995cp} and \cite{An:2012pv}. We have used micrOMEGAs package and adopted the profile given and discussed Refs. in \cite{Zhao:1995cp} and \cite{An:2012pv} to perform our calculations. We thus take a DM halo density profile $\rho(r)$ normalized to a value of $\rho_{\odot}$ at the solar galactic radius of 0.3 GeV/cm$^{3}$. The function $\rho(r)]$ is given by \cite{Zhao:1995cp}
   \begin{equation}
   \label{profile}
     \rho(r) = \rho_{\odot} \left(\frac{R_{\odot}}{r}\right)^{\gamma}\left(\frac{r_{c}^{\alpha}+R_{\odot}^{\alpha}}{r_{c}^{\alpha}+r^{\alpha}}\right)^{\frac{\beta-\gamma}{\alpha}} 
    \end{equation}
We take $R_{\odot}$, the solar galactic radius, to be 8.5 kpc. The other parameters in equation (\ref{profile}) are taken to be $\alpha=1$, $\beta=3$, $\gamma=1$. 

We explore the production of WIMP annihilation $\gamma$-rays obtained with our secluded DM model by following the analysis given in Refs.~\cite{Hooper:2010mq, Daylan:2014rsa}.
In Ref.~\cite{Daylan:2014rsa} it is concluded that the observed $\gamma$-ray spectrum is best fit by WIMPS with a mass $\sim$ 20 -- 50 GeV that annihilate to quarks with a cross section $<\sigma_{ann} v> \ = {\cal{O}}(10^{-26})$ cm$^3$s$^{-1}$. Motivated by that work, we consider here two specific cases for our seculded WIMP models, {\it viz.}, WIMPS with masses of 20 GeV and 32 GeV. With the parameter choices for these models as given and discussed in Section~\ref{density}, our results are compatible with the observed DM density as parametrized by the value $\Omega h^2 \simeq 0.119$, and also with the value for $\langle \sigma_{ann} v\rangle$ required to explain the proposed $\gamma$-ray excess from DM annihilation. We now proceed to demonstrate this more explicitly for the two WIMP mass models that we consider.
\newline \newline
{\it Model 1}: In this first example, we take $m_{\eta} = 20$ GeV, with the parameters as shown in the caption of Figures 2 and 3 chosen to fit the DM svalue for $\Omega h^{2}\approx 0.119$ and the value for $\langle \sigma_{ann} v\rangle \simeq 1.8 \times 10^{-26}$ cm$^{3}$/s. The lifetime of the light mediator, $Z^{\prime}$, must be less than one second in order to guarantee that the $Z^{\prime}$ decays before the epoch of big bang nucleosynthesis~\cite{Serpico:2004nm}. We have chosen $M_V = 0.5$ GeV, which implies that $M_{Z^{\prime}} \simeq 353$ MeV. Given this $M_{Z^{\prime}}$ mass and all the couplings of $Z^{\prime}$ with fermions (see appendix), we then find the $Z^{\prime}$ decay width, $\Gamma_{Z^{\prime}} \simeq 1.76\times 10^{-19}$ GeV. The branching ratios for the $Z^{\prime}$ decay channels are found to be 33.9\% into $u\overline{u}$, 25.4\% into $e^{+}e^{-}$, 24\% in $\mu^{+}\mu^{-}$, 8.48\% into $d\overline{d}$ and 8.18\% into $s\overline{s}$.

In this case, the annihilation $\gamma$-rays primarily result from the decay of the $Z^{\prime}$ into light quarks $u\overline{u},d\overline{d}$, $s\overline{s}$, which accounts for 50.6\% of the decay width. It is shown in Ref.~\cite{Daylan:2014rsa} that the annihilations of
DM particles of mass between $\sim$ 18 and $\sim$ 26 GeV into light quarks may significantly account for the $\sim$ 2 -- 3 GeV excess.

Choosing the set of parameters given by $g_{\eta} = 0.088$, $m_{\eta} = 20$ GeV, $M_{Z^{\prime}}/m_{\eta}\rightarrow 0$, $g_{VB} = 8\times10^{-9}$, and taking into account the constraint for tan~$2\theta_{\alpha}$, designated by $t_{2\alpha}$ in equation (\ref{alpha}), we obtain $s_{\alpha} \simeq -3.8\times 10^{-9}$. In this case, the spin independent cross section of $\eta$ with the nuclei is $\sigma_{SI}^{p}\approx 2.31\times 10^{-46}$ cm$^{2}$ and $\sigma_{SI}^{n}\approx 5.42\times 10^{-57}$ cm$^{2}$. The value of $\sigma_{SI}^{p}$ is of the same order of magnitude as that of the upper limits obtained by the both the LUX (Large Underground Xenon) experiment and the XENON100 experiment for the WIMP mass range that we consider here~\cite{Aprile:2012}. In our calculations, we obtained our parameters using micromegas package~\cite{Belyaev, Belanger:2010pz}.  We note that $\sigma_{SI}^{n}$ and $\sigma_{SI}^{p}$ are dependent on $g_{VB}$ parameter, which in practice connects the dark and visible sectors of the model ({\it cf.}, Model 2 below)\newline \newline

{\it Model 2}: In this model, we take $m_{\eta} = 32$ GeV. We choose the interaction strength~$g_{\eta} = 1.12 \times 10^{-1}$, we take $g_{VB} = 1 \times 10^{-6}$, and taking into account the constraint for $t_{2\alpha}$ in equation (\ref{alpha}), we obtain $s_{\alpha} \simeq -4.7\times 10^{-7}$. Then we increase the mass of the $Z^{\prime}$ mediator to 3.5 GeV. The dominant interaction channel is still $\eta + \overline{\eta} \rightarrow Z^{\prime} + Z^{\prime}$. In this case, we find $\langle \sigma_{ann} v\rangle \simeq 1.88 \times 10^{-26}$ cm$^{3}$s$^{-1}$ and we fit the DM $\Omega h^{2} \simeq 0.119$  from the results shown in Figures~\ref{figure4} and~\ref{figure5}. Our values for the spin independent cross sections of $\eta$ with the nuclei are $\sigma_{SI}^{p}\simeq 6.04 \times 10^{-46}$ and $\sigma_{SI}^{n}\simeq 1.42\times 10^{-52}$ cm$^{2}$.  For $M_{Z^{\prime}} = 3.5$ GeV, the additional $c\overline{c}$ decay channel opens up, being now kinematically allowed. The $Z^{\prime}$ decay width, $\Gamma_{Z^{\prime}}$, will therefore be larger than that for Model 1, in this case $\Gamma_{Z^{\prime}} \simeq 3.74\times 10^{-14}$ GeV. The branching ratios for the $Z^{\prime}$ decay are found to be: 25\% in $u\overline{u}$, 24.9\% in $c\overline{c}$, 18.8\% into $e^{+}e^{-}$, 18.8\% into $\mu^{+}\mu^{-}$, 6.28\% into $d\overline{d}$, 6.28\% into $s\overline{s}$.  Thus, in addition to the other channels of Model 1, the $c\overline{c}$ now also significantly contributes to $\pi^0 \rightarrow \gamma \gamma$ production. For a large, $c\overline{c}$ decay channel it is shown in Ref.~\cite{Daylan:2014rsa} that the annihilations of DM particles of mass between $\sim$ 28 GeV and $\sim$ 36 GeV into $c\overline{c}$ channels may significantly account for the $\sim$ 2 -- 3 GeV excess.

The expression for $Z^{\prime}$ decay width into fermion-antifermion pairs, $f\overline{f}$, is given by
 \begin{eqnarray}\label{larzpf}
  \Gamma_{Z^{\prime}\rightarrow f\overline{f}}&=&\sum_{i}\frac{N_{c}g^{2}}{c_{W}^{2}}\frac{1}{12\pi}\frac{1}{M_{Z^{\prime}}^{2}}\sqrt{M_{Z^{\prime}}^{2}-4m_{i}^{2}}\times\\ \nonumber
  &&[(f_{A}^{i})^{2}(M_{Z^{\prime}}^{2}-4m_{i}^{2})+(f_{V}^{i})^{2}(M_{Z^{\prime}}^{2}+2m_{i}^{2})],
\end{eqnarray}
where the sum runs over quarks and leptons species kinematically allowed. $N_{c}$ is the color number, $m_{i}$ denotes the mass of the fermion and $f_{A,V}^{i}$ denotes axial/vectorial couplings of $Z^{\prime}$ to fermions.
Therefore, in the model considered here, the $Z^{\prime}$ will decay more into up-type quarks than down-type quarks. Even if we increase $M_{Z^{\prime}}$ so that  $M_{Z^{\prime}} > 2m_{b}$, the $b\overline{b}$ channel will still not dominate. We could open this channel kinematically, but this would require a larger mass for $\eta$ in order to keep the primary annihilation channel as to be $\eta + \overline{\eta} \rightarrow Z^{\prime} + Z^{\prime}$ in the mass range $m_{\eta} \gg  M_{Z^{\prime}}$.

If, for example, we take the same set of parameters for {\it Model 2}, only  increasing  $ M_{Z^{\prime}}\sim 9.28$ GeV, so that $Z^{\prime}\rightarrow b\overline{b}$ is now opened, than the branching ratios of $Z^{\prime}$ would be 20.1\% in $u\overline{u}$, 20.1\% in $c\overline{c}$, 15\% into $e^{+}e^{-}$, 15\% into $\mu^{+}\mu^{-}$, 14.9\% in $\tau^{+}\tau^{-}$,  5.15\% into $d\overline{d}$, 5.15\% into $s\overline{s}$ and only 4.57\% into $b\overline{b}$. One of the main phenemological differences between this model and the other ones presented in the literature is that, as can be seen in this example, $Z^{\prime}\rightarrow b\overline{b}$ will not the most probable decay channel. The detailed expression for $f_{A,V}^{i}$ can be found in the appendix.

Considering the values for $g_{\eta}$ chosen for model 2, for instance, if $M_{Z^{\prime}} > 2 m_{\eta}$, the $Z^{\prime}$ decays 100\% into $\eta\overline{\eta}$. However, smaller values for $g_{\eta}$ can suppress this decay and change this ratio. The expression for $Z^{\prime}\rightarrow\eta\overline{\eta}$ is given by

\begin{equation}\label{larzpeta}
  \Gamma_{Z^{\prime}\rightarrow \eta\overline{\eta}}=\frac{1}{12\pi}\frac{1}{M_{Z^{\prime}}^{2}}\frac{g_{\eta}^{2}c_{\alpha}^{2}}{(1-g_{VB}^{2})}\sqrt{M_{Z^{\prime}}^{2}-4m_{\eta}^{2}}(M_{Z^{\prime}}^2+2m_{\eta}^{2}).
\end{equation}

\begin{figure}
  \centering
  \includegraphics[width=7.7cm]{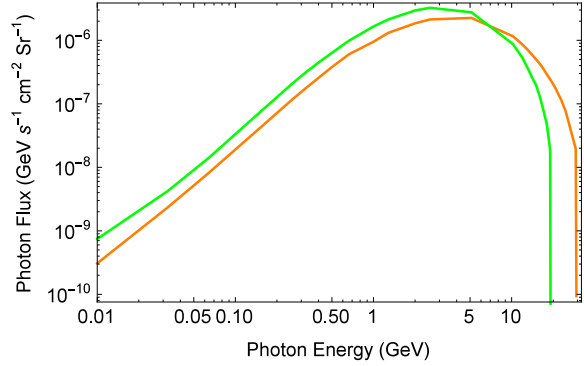}\\
  \caption{Predictions of the excess $\gamma$-ray flux from WIMP annihilation for our two models as described in the text (green: {\it Model 1}, orange: {\it Model 2}). The upper limits of the energy spectrum are determined by the WIMP masses since annihilation occurs near rest. }
  \label{figure7}
\end{figure}

Figure \ref{figure7} shows the $\gamma$-ray flux as a function of the energy of the photon for the specific models that we considered. As with most previous work, we neglect a possible secondary contribution from Compton scattering of the electrons and positrons produced in the annihilations,
although it has been suggested that this process may contribute to the $\gamma$-ray excess~\cite{Abaz15}.
 
\section{Conclusions}
\label{concl}

We have proposed a model in which a Proca-Wentzel (PW) field consisting of $\sim$ MeV--multi-GeV mass particles are mediators of secluded fermionic DM interactions. From the low-energy phenomenology it is known that electroweak precision data such as the mass of $W$ boson, the decay width of $Z$ boson, and some asymmetries, can constrain such a model. For our secluded DM model, the constraints on $\xi/\sqrt{|1-M_{Z}^{\prime 2}/M_{Z}^{2}|}$, where $\xi=g_{VB}/\sqrt{1-g^2_{VB}}$, are well satisfied. Indeed, they are much smaller than ${\cal{O}}(10^{-2})$~\cite{Gopalakrishna:2008dv}. Also, the mediator of DM interactions will not induce flavor-changing neutral current processes.

One important  difference between our model and other ones  presented in the literature is the absence of an extra singlet scalar. Thus, there is no equivalent to a "Higgs scalar"  in the PW Lagrangian. This distinguishing characteristic  may become important if no $h \rightarrow invisible$ width is observed at LHC; such an result could be a clear signature of the model.

The model that we have explored has characteristics in common with the model presented in \cite{Pospelov:2007mp}. Our results agree with theirs in the limit $M_{Z^{\prime}}/m_{\eta} \rightarrow 0$. As those authors point out, there are interesting and testable signatures for secluded DM models. We now summarize various empirical implications of our secluded DM model, based on the results of our calculations. 

\subsection{Indirect tests: Gamma-ray Excess}

We have here explored one of the testable signatures of our secluded WIMP models in quantitative detail. We have considered the $\gamma$-rays that would be produced as a result of cosmic annihilations of secluded WIMPs of mass of $\sim$ 20 and $\sim$ 32 GeV. In particular, we have shown that the secluded DM model proposed here can potentially explain both the flux and spectrum of an apparent 2 -- 3 GeV energy $\gamma$-ray excess in the galactic center region, this being an excess over that expected by taking account of other galactic $\gamma$-ray production processes. Such an excess has been inferred from an analysis of {\it Fermi}-LAT Gamma-ray Space Telescope data~\cite{Hooper:2010mq}. Our results are also consistent with a putative weak $\sim $(2--3)$~\sigma$ $\gamma$-ray signal claimed for the dwarf galaxy Reticulum 2~\cite{Geringer-Sameth:2015} and with the conservative constraints derived from {\it Fermi} Gamma-Ray Space Telescope observations in Refs.~\cite{Massari:2015}. 

Unfortunately, owing to the uncertainties in the empirical derivation of a $\gamma$-ray excess,
it is difficult to distinguish between the $\gamma$-ray spectra obtained for our WIMP models and those that generally follow the analysis of Ref.~\cite{Daylan:2014rsa}. However, we note that
most other DM interpretations of the galactic center excess stress the $b\overline{b}$
channel, whereas in our {\it Model 1} light quarks and leptons dominate in producing the DM annihilation $\gamma$-rays while in our {\it Model 2} the $c\overline{c}$ channels dominate over the $b\overline{b}$ channel. This is a predicted phenomenological difference between our model and other models. At present, owing to the observational uncertainties in determining the DM "signal" over the other processes contributing to the $\gamma$-ray "background" in the direction of the galactic center, a definitive test of this difference is difficult.

We note that the DM interpretation of the "excess $\gamma$-rays from the galactic center region is subject to some caveats. We again note that there are other interpretations of either all or
part of this excess~\cite{Abazajian:2011}. Also, using a new calibration called {\it Pass-8}, the {\it Fermi} collaboration has very recently argued for upper limits on the cross section for the production of $\gamma$-rays from DM in dwarf galaxies that is in tension with the DM interpretation of the galactic center excess~\cite{Wood:2015}. However, this involves modeling of the distribution of the square of the density of annihilating particles along the line-of-sight in these galaxies~\cite{Stecker:1978}, now referred to as the "J-factor"~\cite{Bergstrom:1998}. These models, in turn, rely on velocity dispersion measurements~\cite{Martinez:2015} that do not distinguish between the gravitational effects of DM and faint stars.

\subsection{Indirect tests: Cosmic Ray Antiprotons}

The AMS-02 experiment has obtained detailed data on the cosmic ray antiproton spectrum
in the vicinity of the Earth \cite{AMS}. The implications of these results for DM annihilation models involve various uncertainties such as those of galactic cosmic-ray propagation, cross sections for interstellar $\bar{p}$ production by cosmic rays, and solar modulation. The authors of Ref. \cite{giesen15} have concluded that these data provide no unambiguous evidence for a significant excess above that expected from cosmic ray interactions with interstellar gas. Should $\bar{p}$ production by DM annihilation be absent, we wish to point out that this would favor our {\it Model 1} with a value of $M_{Z^{\prime}}$ below the kinematic threshold for $\bar{p}$ production from
$Z^{\prime}$ decay. On the other hand, $\bar{p}$ production is kinematically allowed in our {\it Model 2}. Thus, should future observations definitively rule out an observable $\bar{p}$ component from DM annihilation, this would support a secluded WIMP model for explaining the galactic center $\gamma$-ray excess.
 
\subsection{Direct Tests}

For both of the secluded WIMP model parameters that we consider here, our calculated values for the cross sections for WIMP-proton scattering are consistent with the present experimental upper limits obtained by the XENON100 and LUX experiments~\cite{Aprile:2012}. The WIMP-neutron cross sections are suppressed. It is anticipated that by 2020 liquid xenon detectors will have the capability to measure spin independent cross sections as low as $\cal{O}$$(10^{-48})$ cm$^2$~\cite{Feng:2014}. Should future laboratory results yield a very small constraint on the WIMP elastic scattering cross section, the DM annihilation hypothesis for explaining the $\gamma$-ray excess from the galactic center region would then favor a secluded WIMP model for the dark matter.
\newline

\section*{acknowledgments}

We thank Julian Heeck and Matthew Wood for helpful discussions. E.C.F.S.F. thanks NASA Goddard Space Flight Center for its hospitality during the preparation of this paper and FAPESP for full support under contracts numbers 14/05505-6 and 11/21945-8. VP thanks to to CNPq for partial support.

\appendix

\section{Interactions and vertices of the model}
\label{sec:int}
Couplings of $Z$ and $Z^{\prime}$ to SM fermions:

\begin{eqnarray}
\overline{\psi}\psi Z: \frac{ig}{c_{W}} c_{\alpha}(1-s_{W}\xi t_{\alpha})\left[T^{3}_{L}-\frac{s_{W}^{2}(1-\xi t_{\alpha}/s_{W})Q}{(1-s_{W}\xi t_{\alpha})} \right], \nonumber \\
\overline{\psi}\psi Z^{\prime}: -\frac{ig}{c_{W}} c_{\alpha}(t_{\alpha}+s_{W}\xi )\left[T^{3}_{L}-\frac{s_{W}^{2}(t_{\alpha}+\xi /s_{W})Q}{(t_{\alpha}+s_{W}\xi )} \right].
 \label{a1}
\end{eqnarray}

Triple gauge bosons couplings:
Comparing to the SM couplings denoted by $R$, they will be:
\begin{eqnarray}
  R_{AW^{+}W^{-}}=1, \nonumber \\
  R_{ZW^{+}W^{-}}=c_{\alpha}, \nonumber \\
  R_{Z^{\prime}W^{+}W^{-}}=-s_{\alpha}.
\end{eqnarray}

Higgs Couplings:
\begin{eqnarray}
   h ff: -i\frac{m_{f}}{{\rm {v_h}}}, \\\nonumber
   h WW: 2i\frac{m_{W}^2}{{\rm {v_h}}}, \\\nonumber
   h Z Z: 2i\frac{m_{Z}^2}{{\rm {v_h}}}(c_{\alpha}-\xi s_{W}s_{\alpha})^2, \\\nonumber
  h Z^{\prime}Z^{\prime}: 2i\frac{m_{Z}^2}{{\rm {v_h}}}(s_{\alpha}+\xi s_{W}c_{\alpha})^2, \\\nonumber
   h Z Z^{\prime}: 2i\frac{m_{Z}^2}{{\rm {v_h}}}[(c_{\alpha}-s_{W}\xi s_{\alpha})(s_{\alpha}+s_{W}\xi c_{\alpha})].
\end{eqnarray}

Coupling of Dirac Fermion and $Z$ and $Z^{\prime}$ gauge bosons:
\begin{eqnarray}
  \overline{\eta}\gamma^{\mu} \eta\, Z: ig_{\eta}\frac{1}{\sqrt{1-g_{VB}^{2}}}s_{\alpha},\\
  \overline{\eta}\gamma^{\mu} \eta\, Z^{\prime}: ig_{\eta}\frac{1}{\sqrt{1-g_{VB}^{2}}}c_{\alpha}.
\end{eqnarray}

The interactions of $Z^ \prime$ in Eq.~ (\ref{a1}) are written in a simplified form. In fact, in order to calculate the $Z^\prime$ decay width we have used couplings in the form of
\begin{equation}\label{lnc}
  \mathcal{L_{NC}}=\frac{g}{2c_{W}}\sum_{i}[\overline{\psi}_{i}\gamma^{\mu}(g_{V}^{i}-g_{A}^{i}\gamma^{5})\psi_{i}Z_{\mu}+\overline{\psi}_{i}\gamma^{\mu}(f_{V}^{i}-f_{A}^{i}\gamma^{5})\psi_{i}Z_{\mu}^{\prime}],
\end{equation}

where $g_{V}^{i}$, $g_{A}^{i}$ denote respectively the vectorial and vector-axial coupling of $Z$ boson with fermions and $f_{V}^{i}$, $f_{A}^{i}$ denote  these couplings but now for $Z^{\prime}$ gauge boson.
The expressions which relate vectorial and vector axial couplings to left and right-handed couplings are given in Eq. \ref{acp2}

\begin{eqnarray}
  g_{A}^{i} &=& \frac{1}{2}(g_{L}^{i}-g_{R}^{i}), \\
  g_{V}^{i} &=& \frac{1}{2}(g_{L}^{i}+g_{R}^{i}),\\
  f_{A}^{i} &=& \frac{1}{2}(f_{L}^{i}-f_{R}^{i}), \\
  f_{V}^{i} &=& \frac{1}{2}(f_{L}^{i}+f_{R}^{i}).
\end{eqnarray}\label{acp2}

The detailed left and right-handed couplings of $Z^{\prime}$ to fermions are given below.
\begin{eqnarray}
  f_{L}^{u} &=& c_{\alpha}(t_{\alpha} + \xi s_{W})\left[T_{3}^{\nu} -\frac{(t_{\alpha} + \xi/s_{W})}{(t_{\alpha} + \xi s_{W})}s_{W}^{2}Q_{u}\right] ,\\
  f_{R}^{u} &=& c_{\alpha}(t_{\alpha} + \xi s_{W})\left[-\frac{(t_{\alpha} + \xi/s_{W})}{(t_{\alpha} + \xi s_{W})}s_{W}^{2}Q_{u} \right], \\
  f_{L}^{d} &=& c_{\alpha}(t_{\alpha} + \xi s_{W})\left[T_{3}^{e} -\frac{(t_{\alpha} + \xi/s_{W})}{(t_{\alpha} + \xi s_{W})}s_{W}^{2}Q_{d}\right],\\
  f_{R}^{d} &=& c_{\alpha}(t_{\alpha} + \xi s_{W})\left[-\frac{(t_{\alpha} + \xi/s_{W})}{(t_{\alpha} + \xi s_{W})}s_{W}^{2}Q_{d} \right], \\
  f_{L}^{e} &=& c_{\alpha}(t_{\alpha} + \xi s_{W})\left[T_{3}^{e} - \frac{(t_{\alpha} + \xi/s_{W})}{(t_{\alpha} + \xi s_{W})}s_{W}^{2}Q_{e}\right],\\
  f_{R}^{e} &=& c_{\alpha}(t_{\alpha} + \xi s_{W})\left[-\frac{(t_{\alpha} + \xi/s_{W})}{(t_{\alpha} + \xi s_{W})}s_{W}^{2}Q_{e}\right], \\
  f_{L}^{\nu} &=& c_{\alpha}(t_{\alpha} + \xi s_{W})T_{3}^{\nu},\\
  f_{R}^{\nu} &=& 0,
\end{eqnarray}
where $Q_{i}$ denotes the charge of fermion, $T_{3}^{e}=-1/2$, $T_{3}^{\nu}=1/2$.

For completeness we write here the right and left-handed couplings of $Z$ to fermions.

\begin{eqnarray}
  g_{L}^{u} &=& [c_{\alpha}(1 - s_{W}t_{\alpha}\xi)]\left[T_{3}^{\nu} - \frac{(1 - t_{\alpha}\xi/s_{W})}{(1 - s_{W}t_{\alpha}\xi)}s_{W}^{2}Q_{u} \right], \\
  g_{R}^{u} &=& [c_{\alpha}(1 - s_{W}t_{\alpha}\xi)]\left[- \frac{(1 - t_{\alpha} \xi/s_{W})}{(1 - s_{W}t_{\alpha}\xi)}s_{W}^{2}Q_{u}\right],\\
  g_{L}^{d} &=& [c_{\alpha}(1 - s_{W}t_{\alpha}\xi)]\left[T_{3}^{e} -\frac{ (1 - t_{\alpha} \xi/s_{W})}{(1 - s_{W}t_{\alpha}\xi)}s_{W}^{2}Q_{d}\right], \\
  g_{R}^{d} &=& [c_{\alpha}(1 - s_{W}t_{\alpha}\xi)]\left[-\frac{(1 -  t_{\alpha}\xi/s_{W})}{(1 - s_{W}t_{\alpha}\xi)}s_{W}^{2}Q_{d} \right], \\
  g_{L}^{e} &=& [c_{\alpha}(1 - s_{W}t_{\alpha}\xi)]\left[T_{3}^{e} -\frac{ (1 - t_{\alpha}\xi/s_{W})}{(1 -s_{W}t_{\alpha}\xi)}s_{W}^{2}Q_{e}\right],\\
  g_{R}^{e} &=& [c_{\alpha}(1 - s_{W}t_{\alpha}\xi)]\left[-\frac{(1 - t_{\alpha}\xi/s_{W})}{(1 - s_{W}t_{\alpha}\xi)}s_{W}^{2}Q_{e}\right],\\
  g_{L}^{\nu} &=& (c_{\alpha}(1 - s_{W}t_{\alpha}\xi))T_{3}^{\nu}, \\
  g_{R}^{\nu} &=& 0.
\end{eqnarray}

\end{document}